\documentclass[10pt,twocolumn,letterpaper]{article}

\usepackage[pagenumbers]{cvpr} 

\definecolor{cvprblue}{rgb}{0.21,0.49,0.74}
\usepackage[pagebackref,breaklinks,colorlinks,allcolors=cvprblue]{hyperref}

\title{Neural Field-Based 3D Surface Reconstruction of Microstructures from Multi-Detector Signals in Scanning Electron Microscopy}

\author{
Shuo Chen$^1$ \and
Yijin Li$^2$ \and
Xi Zheng$^3$ \and
Guofeng Zhang$^{1}$\footnotemark[1] \and
{ $^1$State Key Lab of CAD\&CG, Zhejiang University} \and
{ $^2$Alibaba Group} \and
{ $^3$Analysis Center of Agriculture, Life and Environment Sciences, Zhejiang University}\\
\textnormal{ \tt\footnotesize chenshuo.eric@zju.edu.cn, liyijin.lyj@alibaba-inc.com, xzheng@zju.edu.cn, zhangguofeng@zju.edu.cn}
}

\begin{document}


\twocolumn[{%
\renewcommand\twocolumn[1][]{#1}%
\maketitle
\vspace{-2em}
\centering
\includegraphics[width=0.86\linewidth]{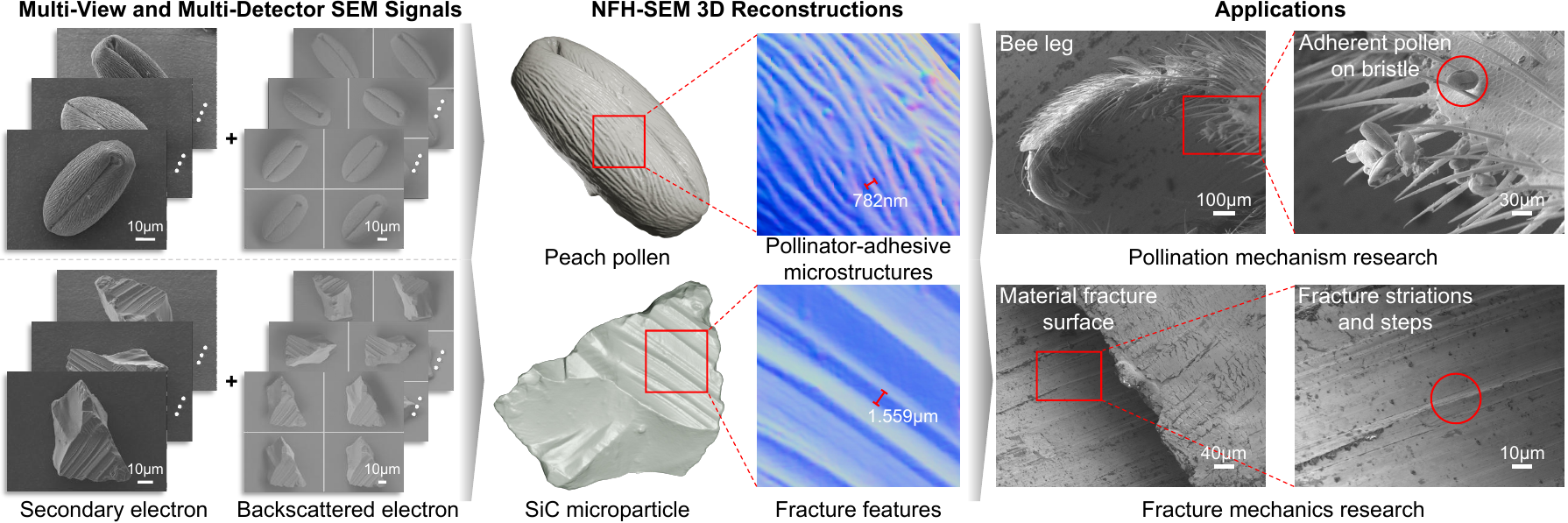}
\vspace{-0.8em}
\captionof{figure}{
\textbf{NFH-SEM} reconstructs detailed microstructures with important functional roles across diverse materials, from pollen textures that enable attachment to pollinators to fracture patterns associated with crack propagation in solids, supporting broad scientific research.
}
\vspace{0.5em}
\label{fig:teaser}
}]

\renewcommand{\thefootnote}{\fnsymbol{footnote}}
\footnotetext[1]{Corresponding author.}
\renewcommand{\thefootnote}{\arabic{footnote}}

\begin{abstract}
The 3D characterization of microstructures is crucial for understanding and designing functional materials. However, the scanning electron microscope (SEM), widely used in scientific research, captures only 2D electron intensity distributions. Existing SEM 3D reconstruction methods struggle with textureless regions, shadowing artifacts, and calibration dependencies, whereas advanced learning-based approaches fail to generalize to microscopic SEM domains due to the lack of physical priors and domain-specific data. We introduce NFH-SEM, a neural field-based hybrid framework that reconstructs high-fidelity 3D surfaces from multi-view, multi-detector SEM images. NFH-SEM integrates coarse multi-view geometry with photometric stereo cues from detector signals through a continuous neural field, incorporating a learnable forward model that embeds SEM imaging physics for self-calibrated, shadow-robust reconstruction. NFH-SEM achieves precise recovery across diverse specimens, revealing 478 nm layered features in two-photon lithography samples, 782 nm surface textures on pollen grains, and 1.559 $\mu$m fracture steps on silicon carbide particles, demonstrating its accuracy and broad applicability. Our code and real-world dataset are available at \href{https://github.com/zju3dv/NFH-SEM}{https://github.com/zju3dv/NFH-SEM}.
\vspace{-2em}
\end{abstract}    
\section{Introduction}
\label{sec:intro}

The 3D characterization of microstructures is particularly important, as surface morphology dictates the functional properties of materials and biological specimens.
As in ~\cref{fig:teaser}, detailed 3D reconstruction reveals the pollen microstructures that enable adhesion to insect pollinators, providing insights valuable for both understanding pollination mechanisms~\cite{peach2} and designing biomimetic surfaces~\cite{pollen_biomimetic_design}.
The scanning electron microscope (SEM)~\cite{reimer2000scanning} is a powerful imaging instrument that uses a focused electron beam to scan sample surfaces, producing high-resolution and large depth-of-field images at the micro- to nanoscales.
It plays a crucial role in materials science~\cite{SEM_Semiconductor}, biology~\cite{nature_life_science}, and industrial manufacturing~\cite{nature_additive_manufacturing}. However, SEM images inherently represent only 2D intensity distributions of scattered electrons captured by detectors, which do not directly reveal the underlying 3D surface geometry.
Consequently, 3D reconstruction from 2D SEM images has emerged as a critical requirement for advancing scientific understanding.

Existing SEM 3D reconstruction techniques can be broadly categorized into multi-view and single-view approaches. Multi-view methods~\cite{3DSEM, SFM54, SFM33} follow the conventional Structure-from-Motion (SfM) and Multi-View Stereo (MVS) pipelines, but often fail on microstructures with weak textures or repetitive patterns. In contrast, single-view methods~\cite{PS14,PS31,PS47} leverage Photometric Stereo (PS) principles to estimate dense surface normals based on electron-scattering behavior and multi-detector signals. However, they require detector calibration with reference samples and are highly sensitive to shadowing artifacts~\cite{PS47}, leading to distorted geometry on discontinuous surfaces.
Some hybrid approaches~\cite{Hybrid_3DSEM1} attempt to combine their advantages, using multi-view methods to initialize coarse geometry and single-view refinement for fine details. However, these methods remain preliminary, inheriting the calibration and shadowing problems of single-view methods, and relying on 2D height-map representations that cannot capture the complete geometry of complex microstructures.

Recent advances in learning-based 3D reconstruction from macroscopic RGB imagery have achieved remarkable success.
However, these methods perform poorly when applied to SEM data.
One major limitation lies in the scarcity of large-scale multi-view SEM datasets, due to the expensive and intricate sample preparation and imaging procedures. 
As a result, models~\cite{mvsnet, cascade_mvs, transmvsnet} trained on natural RGB data, including recent feed-forward methods~\cite{vggt, mapanything}, fail to generalize to microscopic SEM scenes because of the severe domain gap in both scale and appearance.
Even methods that do not rely on large-scale pretraining, such as neural field~\cite{Neus, volsdf, unisurf} and 3D Gaussian Splatting (GS)-based~\cite{2dgs,pgsr} surface reconstruction approaches, perform poorly in the SEM domain because they are designed for optical RGB rendering and fail to capture the geometric cues encoded in multi-detector SEM signals.
Collectively, these limitations restrict the applicability of learning-based frameworks for accurate microscale 3D reconstruction.

To address these challenges, we propose NFH-SEM, a \textbf{N}eural \textbf{F}ield-based \textbf{H}ybrid framework for high-fidelity 3D reconstruction of complex microstructures from multi-view, multi-detector SEM images.
Our key insight is to embed the model of electron scattering and detector response directly into the neural field optimization, enabling NFH-SEM to overcome the limitations of existing SEM reconstruction methods and to achieve parameter self-calibration, automatic shadow separation, and accurate surface recovery.
Specifically, NFH-SEM replaces the simplified 3D representations used in prior works with a continuous neural field that fuses coarse multi-view geometry with photometric stereo cues extracted from multi-detector SEM signals.
The physics of SEM imaging is formulated as a learnable forward model that is jointly optimized with the neural field.
This design enables effective extraction of geometric information from SEM signals and self-calibration of detector parameters without requiring reference samples.
Furthermore, an iterative shadow separation strategy is integrated into the optimization, ensuring reliable reconstruction even under severe SEM shadowing effects.
To evaluate the performance of NFH-SEM, we establish the first multi-view and multi-detector real-world SEM dataset for 3D microstructure reconstruction, covering two-photon lithography (TPL) microstructures, peach pollen, and silicon carbide (SiC) surfaces.
This dataset not only lowers the barrier for researchers to access SEM data but also provides a systematic data collection protocol for further extension.
In our experiments, NFH-SEM consistently delivers detail-rich and accurate reconstructions across diverse samples, demonstrating strong generalization and practical applicability.
Our main contributions are summarized as follows:
\begin{itemize}
\item A neural field-based hybrid 3D SEM framework that achieves accurate modeling of intricate microstructures.
\item A learnable SEM forward model that extracts geometric information and enables self-calibration of the detector.
\item An iterative shadow separation strategy in the neural field optimization for shadow-robust SEM reconstruction.
\item A comprehensive evaluation of NFH-SEM on real and simulated datasets across diverse microstructures.
\end{itemize}

\begin{figure*}[ht] 
    \centering
    \includegraphics[width=0.9\linewidth]{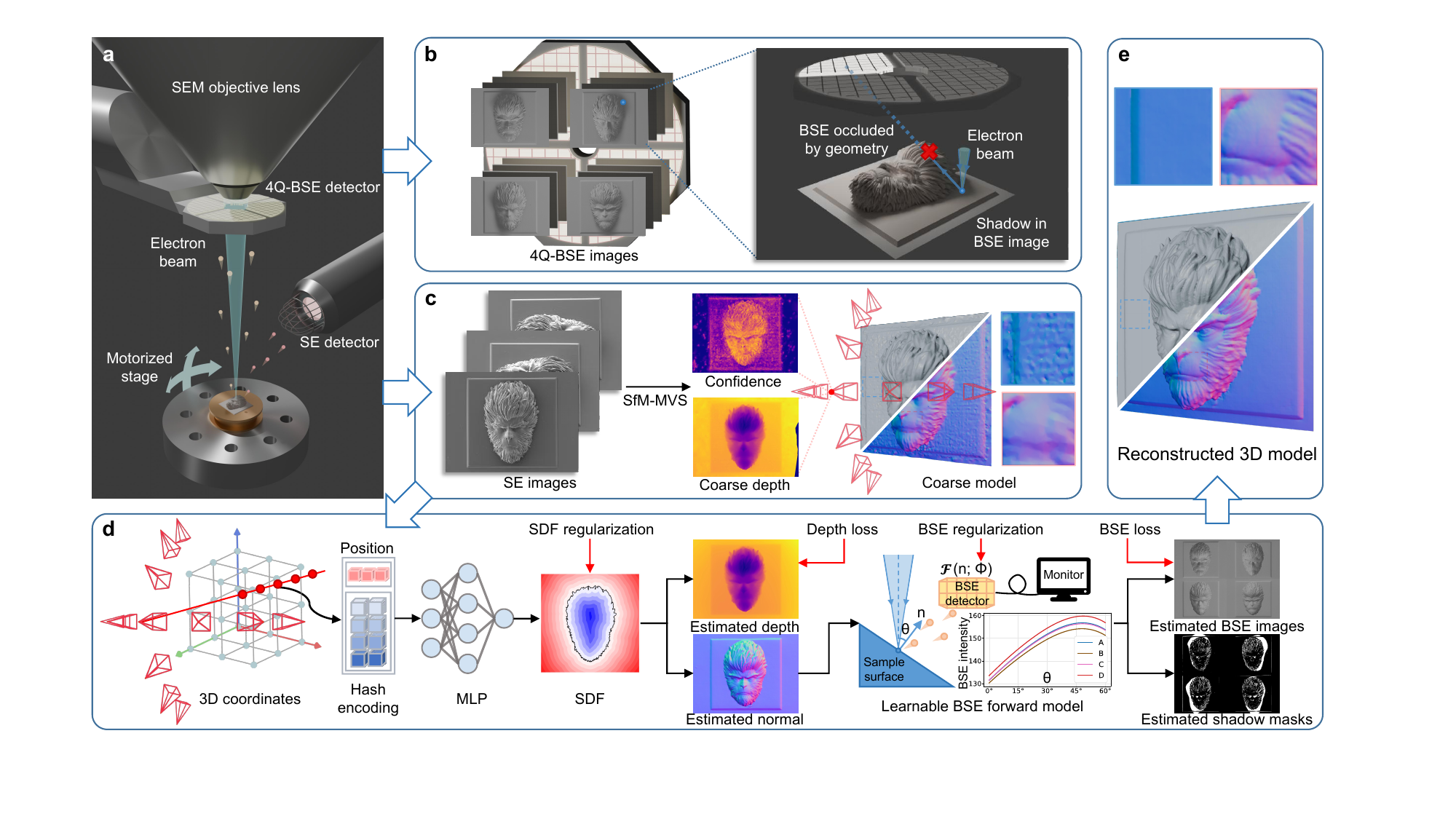}
    \vspace{-0.5em}
    \caption{ \textbf{Workflow of NFH-SEM.} 
    \textbf{(a)} Multi-view and multi-detector SEM scanning of a sample mounted on a motorized stage.
    \textbf{(b)} 4Q-BSE images provide directional illumination with shadows caused by geometric occlusion.
    \textbf{(c)} Multi-view reconstruction initialization.
    \textbf{(d)} The posed depth maps and 4Q-BSE images jointly supervise an SDF-based neural field. A learnable BSE forward model is self-calibrated during training.
    \textbf{(e)} The reconstructed surface extracted from the neural field exhibits high geometric fidelity and rich surface details.
    }
    \label{fig:Pipeline}
    \vspace{-1em}
\end{figure*}

\section{Related Work}
\label{sec:related_work}
\textbf{Conventional SEM 3D surface reconstruction.} 
Multi-view methods~\cite{3DSEM, SFM32, SFM33, CE_SFM, SFM54, SE_Agisoft} estimate surface geometry from a sequence of multi-view SEM images based on the SfM-MVS photogrammetry pipeline~\cite{SFM2, SFM3, SFM1}.
They can recover global geometry when sufficient texture exists, but often fail on smooth or repetitive structures, common in microscopic samples, where feature correspondences become unreliable.
Single-view methods~\cite{PS14,PS30,PS31,PS35,PS41,PS47,PS_3Q,PS16,PS52} exploit multi-detector SEM configurations to estimate dense surface normals using PS techniques~\cite{PS_BKG1, PS_BKG2, PS_BKG3}. Although effective in capturing fine gradients on smooth surfaces, they require detector calibration and are highly sensitive to shadowing artifacts that often distort local geometry.
Hybrid methods~\cite{Hybrid_3DSEM1, Hybrid_3DSEM2, Hybrid_3DSEM3} refine coarse multi-view reconstructions using PS-derived surface gradients. However, their 2D height-map-based fusion provides only a simplified representation that cannot capture complex microstructures. They still require detector calibration and suffer from the shadow-induced gradient errors of single-view methods.
Overall, conventional SEM 3D reconstruction remains limited by restricted geometric representations and insufficient exploitation of SEM signals.

\noindent\textbf{Learning-based 3D surface reconstruction.} 
In recent years, learning-based SfM~\cite{ba_net, deepsfm, droid_slam, dpvo, vggsfm} and MVS~\cite{mvsnet, mvs_rethinking, geomvsnet, transmvsnet, cascade_mvs} methods have shown remarkable progress beyond conventional methods. 
Feed-forward reconstruction models~\cite{dust3r, Mast3r, fast3r, vggt, mapanything} represent a further breakthrough, enabling the direct regression of camera parameters and 3D geometry from multi-view images in a single forward pass.
However, such approaches rely heavily on abundant and diverse training data, which are often unavailable in microscopic imaging scenarios.
In contrast, per-scene optimization methods based on neural scene representations eliminate the need for large-scale pretraining.
Neural fields~\cite{Neural_survey, Nerf} represent the geometry and physical properties~\cite{Micro_neural_field1, Micro_neural_field2, MVN-AFM, OpticFusion} of 3D space using coordinate-based neural networks, and have recently demonstrated remarkable performance in 3D reconstruction~\cite{Neus, volsdf, unisurf, neuralangelo, neus2} by modeling surfaces as continuous Signed Distance Functions (SDFs).
3DGS~\cite{3dgs, 3dgs_survey} provides an alternative by modeling scenes as differentiable Gaussian primitives, and its variants~\cite{2dgs, pgsr, dn_splatter, GOF, sugar} incorporate geometric priors to enhance surface quality.
However, these methods are based on macroscopic RGB rendering models and are not suitable for SEM imaging.
Given the scarcity of SEM data, we are motivated to design a neural field representation tailored to the physics of SEM imaging, capable of extracting geometric information from SEM signals.

\section{Method}
\label{sec:method}
We first describe the characteristics of multi-detector SEM signals and review their use in existing SEM 3D reconstruction methods (\cref{sec:preliminary}). We then introduce NFH-SEM, a novel neural field-based hybrid 3D SEM framework, as shown in \cref{fig:Pipeline} and detailed in \cref{sec:pipeline}. Following the hybrid SEM paradigm, our method first initializes the geometry via multi-view dense reconstruction and then refines it using multi-detector cues. NFH-SEM builds upon a neural field that models electron scattering and performs hybrid fusion through a three-stage optimization. Finally, \cref{sec:optimization} outlines the loss functions and optimization details.

\subsection{Preliminary: Characteristics of SEM Signals}
\label{sec:preliminary}
In SEM, a focused electron beam interacting with the sample surface generates two primary types of signals: secondary electrons (SE) and backscattered electrons (BSE), as shown in \cref{fig:Pipeline}a.
The SE signal exhibits a pronounced edge effect~\cite{reimer2000scanning}, where geometric protrusions and edges emit more SEs, producing higher contrast in these regions of SE images.
As a result, multi-view 3D SEM methods~\cite{3DSEM, SFM33, CE_SFM, SFM54} typically employ multi-view SE images for dense 3D reconstruction of microstructures.

\begin{figure}[h] 
    \centering
    \includegraphics[width=0.6\linewidth]{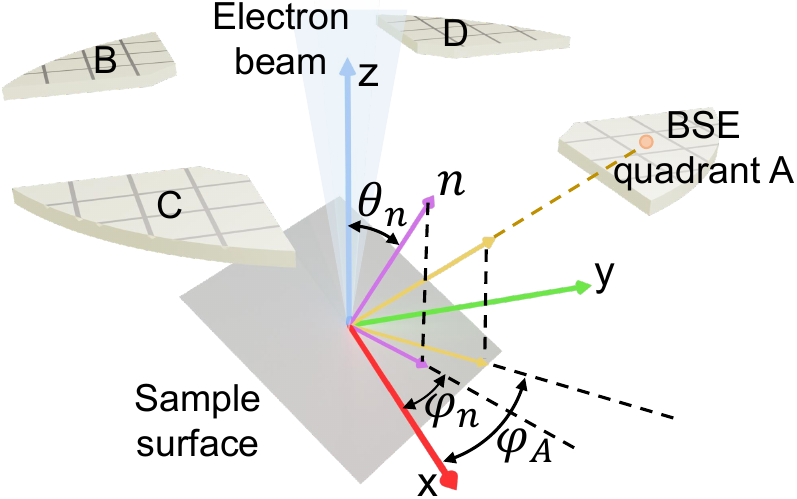}
    \caption{ \textbf{Emission of the BSE signal from the sample surface and its detection by the 4Q-BSE detector.}  
    }
    \label{fig: Function}
    \vspace{-1em}
\end{figure}

In contrast, single-view methods commonly rely on BSE images. The BSE signal is captured by a four-quadrant BSE (4Q-BSE) detector, which consists of a ring-shaped sensor divided into four quadrants positioned beneath the objective lens (\cref{fig:Pipeline}a). Each quadrant independently captures BSE signals from different directions (\cref{fig:Pipeline}b), forming an image resembling illumination from that direction.
Previous studies~\cite{PS14, PS47} have established that the intensity $I_i$ of BSE signals detected by quadrant $i \in \{A, B, C, D\}$ can be modeled as a function of the surface normal $n$, based on the observed characteristics of BSE emission:
\begin{equation}
\label{equ: T}
I_i(n) = \mathbf{R}_0(\theta_n) \left[ d_i \cos(\varphi_i - \varphi_n) \sin(\theta_n) + c_i \cos(\theta_n) \right], \\
\end{equation}
where $\theta$ and $\varphi$ define polar coordinates of $n$ and directions toward four quadrants as in \cref{fig: Function}.
Parameters $c$ and $d$ are related to the BSE detector configuration, electron beam current, and emission coefficient.
$\mathbf{R}_0$ models the emission magnification as a function of the incidence angle $\theta$, and is often defined as the inverse cosine of $\theta$ in prior works:

\begin{equation}
\mathbf{R}_0(\theta) = \sec(\theta).
\end{equation}
This allows the sine and cosine terms in \cref{equ: T} to cancel out, yielding the following expression for the BSE signal:
\begin{equation}
\label{equ: I0}
I_i(n) = d_i \cos(\varphi_i - \varphi_n) \tan(\theta_n) + c_i.
\end{equation}
This expression forms the theoretical foundation of conventional single-view methods~\cite{PS14,PS16,PS31,PS41,PS47}. 
Given the symmetric layout of the 4Q-BSE detector along two orthogonal axes, we define the X-axis as the line connecting quadrants $A$ and $B$, with $\varphi_A = \varphi_B + \pi$.
Under the standard assumption that all quadrants share identical parameters $c$ and $d$, a useful idealization with slight practical deviations~\cite{PS31}, the surface gradient along the X-direction can be derived from the intensity measurements as:
\begin{equation}
\label{equ: gradient}
\frac{\partial z}{\partial x} = \tan \theta_n \cos \varphi_n = \frac{c}{d} \frac{I_A(n) - I_B(n)}{I_A(n) + I_B(n)}.
\end{equation}
The Y-direction gradient is computed analogously from quadrants $C$ and $D$, and the complete surface geometry is then recovered by integrating the estimated gradient field.

However, due to geometric self-occlusion, BSE signals may be blocked from reaching certain quadrants, resulting in dark regions in the corresponding BSE images, known as the SEM shading effect (\cref{fig:Pipeline}b).
These shadows are unrelated to local surface gradients and thus violate the assumptions of prior single-view methods~\cite{PS31, PS35, PS47}, leading to inaccurate gradient estimation.
Additionally, previous approaches require complex calibration using reference samples to determine the unknown parameters $c$ and $d$.

\subsection{NFH-SEM Reconstruction Pipeline}
\label{sec:pipeline}
\noindent\textbf{SEM data acquisition.} 
During data acquisition (\cref{fig:Pipeline}a), the sample is mounted on a motorized stage inside the SEM chamber, and the stage is then systematically tilted and rotated to capture a comprehensive set of views.
At each viewpoint, an in-chamber SE detector and a 4Q-BSE detector, standard configurations in modern SEM systems, simultaneously capture one SE image and four BSE images.

\noindent\textbf{Multi-view reconstruction initialization.} 
Following prior works~\cite{SFM33, CE_SFM, SFM54}, we perform multi-view dense reconstruction on SE image sequences to initialize the hybrid reconstruction process (\cref{fig:Pipeline}c).
This step provides a coarse geometric estimate but is limited in textureless regions where reliable feature correspondences cannot be established.
From the recovered camera parameters, we compute per-view depth maps $z$ and, when available, project surface-level confidence estimates~\cite{Agisoft_Metashape} to generate corresponding confidence maps $w$.
These outputs jointly serve as geometric priors for subsequent neural field optimization.

\noindent\textbf{Network architecture and learnable BSE forward model.} 
In NFH-SEM (\cref{fig:Pipeline}d), the 3D geometry is implicitly modeled as an SDF~\cite{SDF_Loss} parameterized by a multilayer perceptron (MLP)~\cite{Neus}, denoted as $\mathcal{M}$ with learnable parameters $\Theta$. 
Each SEM pixel is back-projected to a viewing ray. 3D points $x$ are sampled along the corresponding ray and encoded via multi-resolution hash encoding~\cite{instant_ngp} before being fed into $\mathcal{M}$ to predict SDF values $s = \mathcal{M}(x; \Theta)$. 
Instead of directly supervising surface gradients using \cref{equ: gradient}  (the w/o BSE-$\mathcal{F}$ setting in ablation studies), we innovatively integrate the BSE signal generation model into the volume rendering pipeline.
Specifically, we model the mapping from predicted surface normals $\widehat{n} = \nabla s$ to 4Q-BSE intensities as a learnable forward model $\mathcal{F}$ with parameters $\Phi$.
Since we observe that the empirical formulation in \cref{equ: I0} does not accurately capture real BSE responses  (the w/o Poly-$\mathbf{R}$ setting in ablation studies), we replace its emission term $\mathbf{R}_0$ with a more flexible fourth-order polynomial:
\begin{equation}
\mathbf{R}(\theta) = 1 + p_1\theta + p_2\theta^2 + p_3\theta^3 + p_4\theta^4.
\end{equation}
Our complete BSE forward model $\mathcal{F}$ is expressed as: 
\begin{equation}
\label{equ: our_forward_model}
\mathcal{F}_i(n) = \mathbf{R}(\theta_n) \left[ d_i \cos(\varphi_i - \varphi_n) \sin(\theta_n) + c_i\cos(\theta_n) \right] + e_i.
\end{equation}
Because the captured BSE images are not raw detector signals, we introduce an additional parameter $e$ to account for intensity shifts due to brightness adjustments. 
As in prior work~\cite{Hybrid_3DSEM1}, we assume a homogeneous electron emission coefficient, which is reasonable for our coated samples and low accelerating voltage conditions~\cite{reimer2000scanning}.
Due to manufacturing and installation tolerances in real SEM systems, the four quadrants of the BSE detector are not perfectly symmetric with identical parameters~\cite{PS31}. To account for this, our forward model assigns independent learnable parameters to each quadrant.
Overall, $\mathcal{F}$ contains 16 learnable parameters, $\Phi = \{c, d, e, p \}$, where $c$, $d$ and $e$ differ across quadrants, and $p$ denotes the four polynomial coefficients.
A loss $\mathcal{L}_{\text{BSE}}$ is then defined between the synthesized BSE images $\mathcal{F}(\widehat{n}; \widehat{\Phi})$ and the captured images $b$, enabling joint optimization of both the neural field $\widehat{\Theta}$ and the estimated forward model $\widehat{\Phi}$ via backpropagation.
These designs adaptively fit $\mathcal{F}$ to real BSE signal distributions, eliminate detector calibration, and effectively embed surface gradient cues of 4Q-BSE images into the neural field (\cref{fig:Pipeline}e).

\noindent\textbf{Iterative BSE shadow separation.} 
We introduce an iterative masking scheme that excludes shadowed regions in BSE images from supervision, thereby further improving reconstruction accuracy.
In our experiments, we observe that regions where $\mathcal{F}$ significantly deviates from the BSE image $b$ closely align with shadowed regions, as these areas cannot be modeled as functions of $\widehat{n}$.
Based on this observation, we iteratively refine binary shadow masks $S$ using the condition
$S= (\left|\mathcal{F}(\widehat{n}; \widehat{\Phi}) - b \right| < \alpha d)$. 
Here, the threshold $\alpha d$ dynamically updates with the parameter $d$ during training, and $\alpha$ is a fixed proportionality factor.
Since the parameter $d$ in \cref{equ: our_forward_model} controls the sensitivity of BSE intensity to the surface normal, larger $d$ values lead to greater intensity deviations for the same geometric perturbation.
Thus, setting the threshold proportional to $d$ prevents the misclassification of geometry-induced variations as shadows.
This strategy establishes a self-reinforcing feedback loop during training. 
Progressively refined shadow masks yield cleaner BSE supervision, enabling more accurate geometry and forward-model estimation, which in turn further improves shadow detection.
Ultimately, this process produces shadow-free synthesized 4Q-BSE images and per-view shadow masks, enabling reliable surface reconstruction even under severe BSE shadowing.

\subsection{Optimization}
\label{sec:optimization}
\noindent\textbf{Coarse geometry supervision.} 
The estimated depth $\widehat{z}$ from the neural field is supervised by a weighted depth loss:
\begin{equation}
\mathcal{L}_{d} = \frac{1}{M}\sum_{j}^{M} w_{j}\left|\widehat{z_j} - z_j\right|,
\end{equation}
where $M$ is the batch size and $j$ indexes individual rays. 
This loss downweights the influence of uncertain regions in multi-view dense reconstruction and provides a coarse geometric prior for initialization.

\noindent\textbf{SDF regularization.} 
A standard constraint~\cite{SDF_Loss} enforces a unit-norm SDF gradient $\nabla{s}$ to maintain a valid SDF:
\begin{equation}
\mathcal{R}_s = \frac{1}{MN} \sum_{j,k}^{M,N} \left( \|
\nabla{s}_{jk}\| - 1\right)^2,
\end{equation}
where $N$ is the number of samples per ray and $k$ is its index.

\noindent\textbf{4Q-BSE photometric stereo supervision.} 
The BSE loss is defined as the mean absolute error (MAE) between the input 4Q-BSE images $b$ and the estimated images from $\mathcal{F}$ applied to the surface normal $\widehat{n}$ inferred from the neural field:
\begin{equation}
\mathcal{L}_{\text{BSE}} = \frac{1}{4M}\sum_{i}^{A,B,C,D}\sum_{j}^{M} S_{ij}\odot\left| \mathcal{F}_i(\widehat{n}_j; \widehat{\Phi})  - b_{ij} \right|,
\end{equation}
where $S$ is the shadow mask, and shadowed pixels are 0.

\noindent\textbf{BSE forward model regularization.} 
The parameters of the four quadrants are inherently close because the BSE detector is fabricated to be symmetric and only subject to minor variations due to manufacturing and alignment disturbances in SEM systems~\cite{PS31}.
To this end, we apply a regularization term $\mathcal{R}_{\Phi}$ to penalize excessive parameter deviations:
\begin{equation}
\mathcal{R}_{\Phi} = \operatorname{Var}(c) + \operatorname{Var}(d) + \operatorname{Var}(e),
\end{equation} 
where $\operatorname{Var}(\cdot)$ denotes the variance computed within each parameter group. 
This encourages quadrant consistency while avoiding suboptimal convergence.

\begin{figure*}[ht] 
    \centering
    \includegraphics[width=0.9\linewidth]{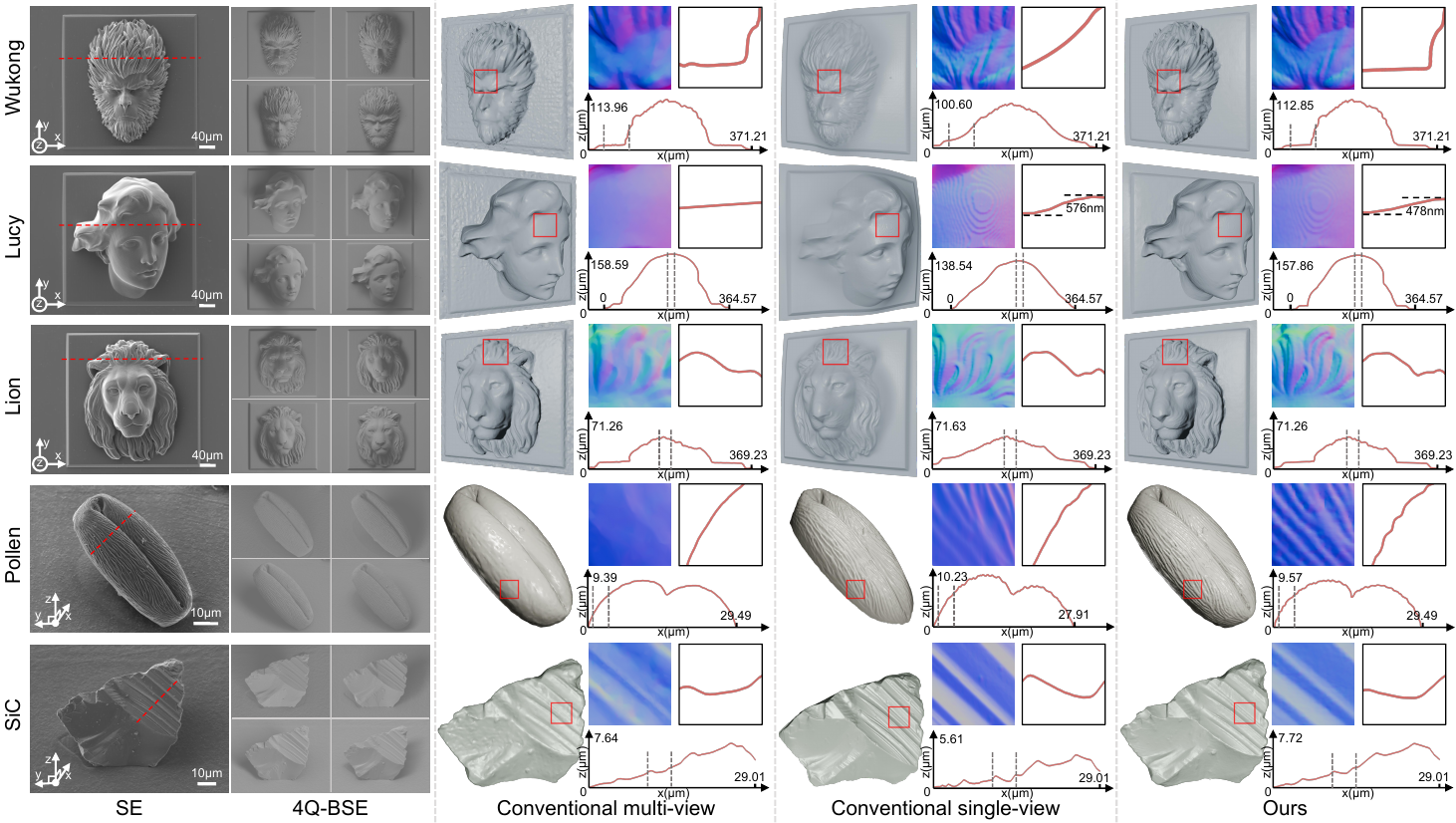}
    \vspace{-1em}
    \caption{ 
    \textbf{Qualitative comparison with conventional SEM 3D reconstruction methods on real-world dataset.} Red boxes highlight enlarged normal maps. Red lines in SE images indicate the cross-sectional profiles, where the zoomed-in segments are marked between vertical gray lines. NFH-SEM faithfully reconstructs the overall geometry and fine surface details, consistent with SEM observations.
    }
    \label{fig:Conventional_Results}
    \vspace{-1em}
\end{figure*}

\noindent\textbf{Three-stage training.}
To effectively integrate hybrid geometric cues into the neural field while maintaining training stability, we adopt a three-stage training strategy.
In the first stage, the neural field is supervised only by coarse depth information through $\mathcal{L}_{d}$ and $\mathcal{R}_s$.
This step initializes the SDF of the neural field with a coarse geometric prior, facilitating subsequent optimization.
In the second stage, we incorporate surface gradient cues from 4Q-BSE images and jointly optimize the model $\mathcal{F}$ by minimizing $\mathcal{L}_{\text{BSE}}$ and $\mathcal{R}_{\Phi}$.
This phase enables the neural field to learn the correlation between surface normals and BSE responses.
Shadow masking is not yet applied at this stage, since the surface normals and the parameters $\Phi$ are still being refined and cannot reliably distinguish shadowed regions.
In the third stage, we activate dynamic shadow masking in $\mathcal{L}_{\text{BSE}}$ to exclude shadow regions from supervision. 
This refinement stage further enhances both the geometric fidelity of the reconstructed microstructures and the accuracy of the model $\mathcal{F}$.
The overall training objective $\mathcal{L}$ is defined as:
\begin{equation}
\label{equ: stages}
\mathcal{L}  =
\begin{cases}
\lambda_1\mathcal{L}_{d} + \lambda_2\mathcal{R}_s,&\text{Stage I}, \\ 
\lambda_1\mathcal{L}_{d} + \lambda_2 \mathcal{R}_s +
\lambda_3\mathcal{L}_{\text{BSE}}(1) + \lambda_4\mathcal{R}_{\Phi},&\text{Stage II}, \\ 
\lambda_1\mathcal{L}_{d} + \lambda_2 \mathcal{R}_s +
\lambda_3\mathcal{L}_{\text{BSE}}(S) +  \lambda_4\mathcal{R}_{\Phi},&\text{Stage III}. \\ 
\end{cases}
\end{equation}
Here, $\mathcal{L}_{\text{BSE}}(1)$ and $\mathcal{L}_{\text{BSE}}(S)$ correspond to the BSE loss computed without and with shadow masking, respectively.

\section{Experiments}
\label{sec:experiments}

\subsection{Implementation Details}
The network is optimized using the Adam optimizer~\cite{adam} with a learning rate of 0.01. 
Training is conducted on a single NVIDIA RTX 4090 GPU.
We perform 1,000 iterations in each training stage.
The entire training process for one microstructure consists of 3,000 iterations and takes only about 2 minutes, demonstrating the high efficiency of NFH-SEM.
After training the neural field, we extract meshes using the Marching Cubes algorithm \cite{Marching_cubes_98} for visualization.
The multi-view initialization step is performed using Agisoft Metashape~\cite{Agisoft_Metashape}, where the surface confidence value of each mesh is the average number of views contributing to its reconstructed vertices. These confidence values are then normalized to the unit range.
Further details are provided in the supplementary material.

\subsection{Evaluation on Real-World Dataset}
\noindent\textbf{NFH-SEM dataset.}
To evaluate the effectiveness of NFH-SEM, we collect the first multi-view and multi-detector SEM dataset, covering a wide range of samples with diverse geometric features. This dataset includes complex TPL-fabricated microstructures, as well as peach pollen and SiC particle surfaces. 
TPL~\cite{TPL_review, TPL_origin_nature, TPL_HB} is a nanoscale additive manufacturing technique based on two-photon absorption-induced photopolymerization, which can 3D-print microstructures with fine features.
Accurate 3D reconstruction of printed structures is crucial for optimizing process parameters and improving fabrication quality.
We fabricate three intricate TPL samples—Wukong, Lucy, and Lion—featuring significant height variations, fine geometric details, strong shadowing in 4Q-BSE images, and large textureless regions, all of which pose major challenges for existing methods.
Pollen grains are representative biological microstructures with complex surface morphologies. 
We select Okubo peach pollen~\cite{okubo_peach}, whose densely packed yet shallow surface textures facilitate adhesion to insect pollinators~\cite{peach2} and contribute to species identification~\cite{pollen_identification}.
The SiC particles are obtained by mechanically fracturing SiC crystals~\cite{SiC_review2}, producing striated and stepped fracture surfaces characteristic of brittle materials.
Accurate 3D characterization of such surfaces is essential for material failure analysis and fracture mechanics research~\cite{fracture1,fracture2,fracture3}.
All experiments are conducted using a ZEISS Gemini 560 SEM system equipped with a field-emission gun.
For each sample, the stage is tilted along two orthogonal axes from -45$^{\circ}$ to +45$^{\circ}$ in 5$^{\circ}$ increments, resulting in 37 distinct viewpoints. 
At each viewpoint, an SE image is captured using an Everhart–Thornley detector, and four BSE images are simultaneously recorded using a 4Q-BSE detector.
All SEM images have a resolution of 1024 $\times$ 768 pixels.
Further details are available in the supplementary material.
\begin{figure*}[ht] 
    \centering
    \includegraphics[width=0.85\linewidth]{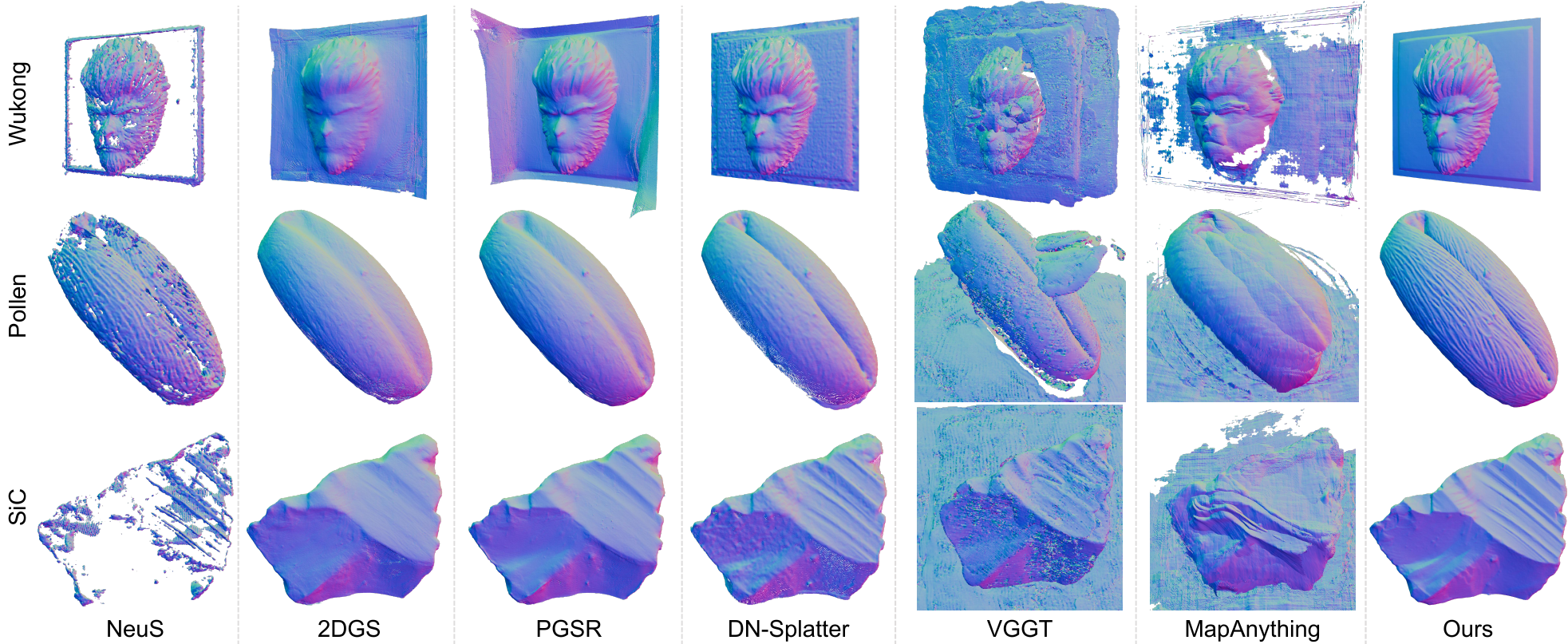}
    \vspace{-0.5em}
    \caption{ \textbf{Qualitative comparison with learning-based 3D reconstruction methods on real-world dataset.} 
    NFH-SEM achieves more accurate reconstructions than approaches that neglect the SEM signal generation model or lack generalization to SEM domains.
    }
    \label{fig:Learning_Results}
    \vspace{-1em}
\end{figure*}

\noindent\textbf{Comparison with conventional SEM 3D reconstruction methods.} We compare NFH-SEM with two representative SEM 3D reconstruction methods.
The first is a multi-view photogrammetry method implemented using Agisoft Metashape~\cite{Agisoft_Metashape}, a standard pipeline widely adopted in prior works~\cite{SFM33, CE_SFM, SFM54, SE_Agisoft} for reconstructing from multi-view SE images.
The second is a single-view method, which estimates surface gradients from 4Q-BSE images~\cite{PS14} following \cref{equ: gradient} and then reconstructs the 3D surface model via global least-squares integration using MATLAB toolbox grad2Surf~\cite{grad2Surf_paper1, grad2Surf_paper2, grad2Surf_paper3}.
In the single-view method, the ratio $c/d$ in \cref{equ: gradient} determines the height scaling of the reconstructed surface.
For a fair comparison, we scale its reconstructed height range to match that of the multi-view baseline.

As shown in \cref{fig:Conventional_Results}, NFH-SEM consistently outperforms conventional baselines in both global geometry and fine-scale surface reconstruction. 
The multi-view method captures only coarse shapes and fails to recover smooth base surfaces or fine details such as hair strands, layered steps, and pollen textures, resulting in over-smoothed or incomplete geometry. 
The single-view method recovers limited surface textures, suffers from global distortions caused by inaccurate gradient estimation in shadowed or steep regions, and struggles with discontinuous height variations.
In contrast, the results of NFH-SEM align closely with SEM observations, faithfully preserving biologically and physically meaningful detail—from adhesive surface textures on pollen grains to the 500 nm printing layer on the Lucy sample (478 nm estimated by NFH-SEM, 576 nm in the single-view result, and not captured in the multi-view result) and the striated steps of fractured materials.
These results collectively demonstrate the robustness and precision of NFH-SEM in capturing real-world SEM microgeometry, effectively overcoming the limitations of existing SEM 3D reconstruction approaches.

\noindent\textbf{Comparison with learning-based surface reconstruction methods.}
We evaluate six representative learning-based surface reconstruction methods using posed multi-view SE images as input.
NeuS~\cite{Neus} serves as an SDF-based neural field baseline.
2DGS~\cite{2dgs} and PGSR~\cite{pgsr} are 3DGS-based approaches initialized from dense point clouds of the multi-view coarse reconstruction, while DN-Splatter~\cite{dn_splatter} further incorporates monocular coarse depth supervision during training.
We also include two advanced feed-forward reconstruction models, VGGT~\cite{vggt} and MapAnything~\cite{mapanything}, to assess their generalization performance on SEM data.

As shown in \cref{fig:Learning_Results}, these learning-based methods fail to accurately reconstruct complex microstructures when directly applied to SEM data. 
The neural field- and 3DGS-based methods treat SEM signals as standard multi-view RGB images, neglecting the underlying SEM imaging physics, which leads to severe surface distortions and incomplete geometry.
Pretrained feed-forward reconstruction networks fail to generalize to SEM domains, producing fragmented surfaces and unstable pose estimates. In contrast, NFH-SEM effectively extracts geometric cues from multi-detector SEM signals, achieving accurate and generalizable surface reconstructions across diverse samples.

\begin{figure*}[ht] 
    \centering
    \includegraphics[width=0.95\linewidth]{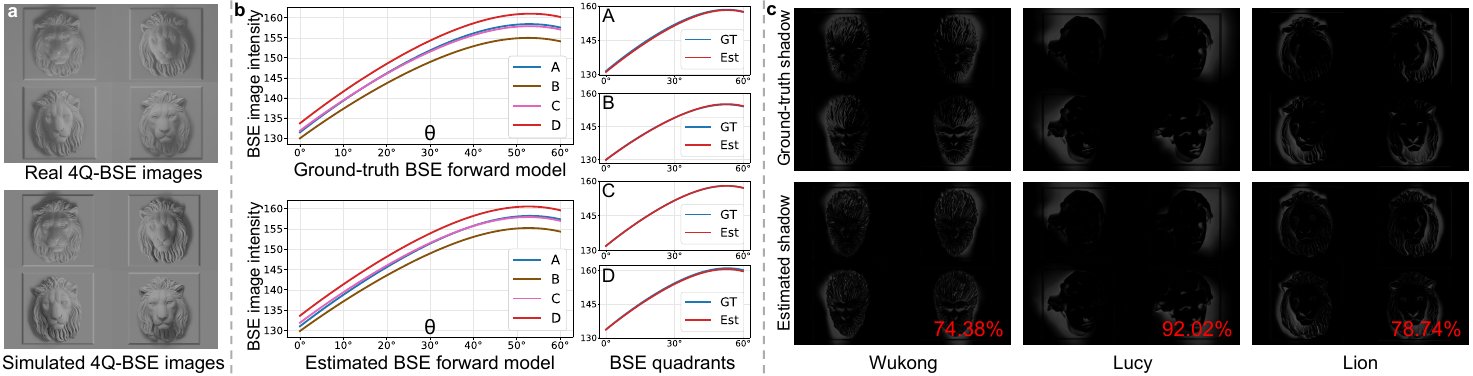}
    \vspace{-0.7em}
    \caption{ \textbf{Simulation results.} 
    \textbf{(a)} Our simulated 4Q-BSE images are highly consistent with real 4Q-BSE observations, demonstrating our accurate modeling of BSE signal formation.
    \textbf{(b)} These curves are obtained by fixing $\varphi$ and the learned parameters in \cref{equ: our_forward_model} to derive the relationship between BSE intensity and $\theta$. The estimated BSE forward model closely matches the ground truth across all detector quadrants. 
    \textbf{(c)} NFH-SEM automatically separates most shadowed regions in the 4Q-BSE images, producing shadow intensity maps that align well with the ground truth. 
    The percentage at the bottom right denotes the shadow detection accuracy (see supplementary material).
    }
    \label{fig:Sim_Results}
    \vspace{-0.5em}
\end{figure*}

\begin{table*}[ht]
  \centering
  \scriptsize
  \setlength{\tabcolsep}{4pt}
  \begin{tabular}{c|cccc|cccc|cccc}
    \toprule
     
      & \multicolumn{4}{c|}{Chamfer distance (nm) $\downarrow$}
      & \multicolumn{4}{c|}{Normal angle error (degrees) $\downarrow$} 
      & \multicolumn{4}{c}{BSE model error (pixel intensity) $\downarrow$} \\
    \cmidrule(lr){2-5}\cmidrule(lr){6-9}\cmidrule(lr){10-13}
      & Wukong & Lucy & Lion & Avg 
      & Wukong & Lucy & Lion & Avg
      & Wukong & Lucy & Lion & Avg \\
    \midrule
    Input coarse model 
      & 37.77 & 16.20 & 21.34 & 25.11 
      & 10.12 & 5.96 & 7.46 & 7.85
      & -- & -- & -- & -- \\
    Single-view PS 
      & 580.26 & 531.76 & 424.65 & 512.22 
      & 11.72 & 15.54 & 11.70 & 12.99
      & -- & -- & -- & -- \\
    Ours (w/o BSE-$\mathcal{F}$)
      & 108.86 & 195.87 & 102.11 & 135.61
      & 7.99 & 7.20 & 7.26 & 7.48
      & -- & -- & -- & -- \\
    Ours (w/o Poly-$\mathbf{R}$)
      & 31.66 & 10.61 & 17.61 & 19.96 
      & 6.28 & 2.45 & 4.30 & 4.34
      & 7.60 & 7.11 & 6.77 & 7.16 \\
    Ours (w/o 4Q-Var)
      & 32.66 & 8.39 & 18.65 & 19.90 
      & 5.95 & 1.85 & 3.94 & 3.91
      & 1.28 & 1.39 & 1.39 & 1.35 \\
    Ours (w/o S-Mask)
      & 38.92 & 20.09 & 29.14 & 29.38
      & 6.28 & 2.29 & 4.50 & 4.36
      & 1.10 & 0.33 & 0.40 & 0.61 \\
    Ours 
      & \textbf{29.81} & \textbf{7.34} & \textbf{15.28} & \textbf{17.48} 
      & \textbf{5.78} & \textbf{1.60} & \textbf{3.71} & \textbf{3.70}
      & \textbf{0.36} & \textbf{0.21} & \textbf{0.25} & \textbf{0.27} \\
    \bottomrule
  \end{tabular}
  \vspace{-0.5em}
  \caption{
  \textbf{Quantitative ablation results on simulated dataset.}
  Results are scaled to the physical dimensions of real TPL microstructures, and chamfer distances are reported in nm. 
  The BSE model error is defined as the MAE between the estimated BSE intensity--$\theta$ mapping curve and the ground-truth curve, measuring how accurately the BSE forward model is learned (see supplementary material).
  Missing values are denoted by “--” because the corresponding methods do not estimate the BSE forward model. Best results are highlighted in bold.
  }
  \vspace{-2em}
  \label{tab:sim_all}
\end{table*}

\subsection{Ablation Study on Simulated Dataset}
\label{sec:evaluation_sim_dataset}
\noindent\textbf{Construction of simulated dataset.}
Obtaining accurate ground-truth geometry at the microscopic scale is highly challenging. 
To quantitatively evaluate NFH-SEM in geometric reconstruction, BSE forward modeling, and shadow separation, we construct a simulated dataset derived from the CAD models used to fabricate TPL microstructures. 
NFH-SEM requires two inputs for reconstruction: multi-view 4Q-BSE images and an initial coarse geometry.
For the 4Q-BSE data, cameras are configured to replicate the viewpoints of real SEM scans. Shadow-free 4Q-BSE images are first generated using the pretrained model $\mathcal{F}$, while quadrant-specific soft shadows are synthesized via ray tracing to simulate realistic illumination and occlusion effects (\cref{fig:Sim_Results}a). 
As in prior multi-view SEM reconstruction studies~\cite{SFM33, 3DSEM, SFM54}, we do not simulate multi-view SE images for quantitative evaluation, due to the absence of an established SE generation model that maps 3D geometry to realistic SE signals.
As a substitute, we synthesize a coarse geometry by perturbing the CAD model with blur and noise to approximate the initialization used in NFH-SEM.
Since we focus on the hybrid step of fusing coarse reconstructions with multi-view 4Q-BSE information, this dataset is suitable for ablation studies of NFH-SEM but not for quantitative evaluation of other multi-view methods. Additional details are provided in the supplementary material.

\noindent\textbf{Ablation settings.}
We design four ablation variants to evaluate key components of NFH-SEM.
\textbf{Direct gradient supervision (w/o BSE-$\mathcal{F}$):} 
Replaces the learnable BSE forward modeling in NFH-SEM with direct supervision of the neural field using single-view surface gradients computed from \cref{equ: gradient}. The loss is the MAE of surface gradients, with the scale ratio $c/d$ jointly optimized during training.
\textbf{Simplified BSE emission model (w/o Poly-$\mathbf{R}$):} Replaces the proposed polynomial reflectance term with the analytical model in \cref{equ: I0} as the BSE forward model, where each quadrant has independent learnable parameters $c$ and $d$.
\textbf{Shared detector parameters (w/o 4Q-Var):} Enforces all quadrants to share a single set of parameters ${c, d, e}$, disregarding inter-quadrant differences in detector response.
\textbf{No shadow masking (w/o S-Mask):} Removes the iterative shadow separation mechanism. Training includes only the first two stages in \cref{equ: stages}, while keeping the total iterations unchanged.

\noindent\textbf{Results analysis.}
As summarized in \cref{tab:sim_all}, the full NFH-SEM achieves the highest overall performance among all variants. 
The single-view baseline exhibits significant geometric errors due to inaccurate gradient integration. Using these gradients directly as supervision further degrades reconstruction quality. Simplified forward models increase BSE modeling errors, which propagate to the neural field and impair geometric accuracy.
In contrast, NFH-SEM accurately learns the BSE forward model across quadrants (\cref{fig:Sim_Results}b), leading to substantial improvements in geometry. 
The shadow separation module achieves a mean shadow detection accuracy of 81.7\% (\cref{fig:Sim_Results}c) and effectively prevents shadow artifacts from corrupting both geometry and the BSE model.
Overall, these results demonstrate that each component of NFH-SEM is essential for achieving precise 3D reconstruction from SEM data.
Metric definitions and additional qualitative ablations on real TPL data are provided in the supplementary material.  

\section{Conclusion}
\label{sec:conclusion}
We present NFH-SEM, a neural field-based hybrid SEM 3D reconstruction framework.
By jointly optimizing a BSE forward model and a neural representation, it achieves self-calibrated and shadow-robust recovery of intricate microstructures.
Experiments on both real-world and simulated datasets demonstrate accurate reconstruction of global geometry and fine-scale details across diverse materials, outperforming conventional and learning-based methods.

Some SEM-specific effects remain unmodeled, including charging effects on low-conductivity samples that can cause pixel drift and unstable multi-view alignment~\cite{reimer2000scanning}, as well as material-dependent emission variability that may confound material contrast with geometric variation.
In addition, extreme geometries with severe occlusion (e.g., microporous materials) may be fully shadowed across all BSE quadrants, limiting recoverable geometric information from SEM signals.
We hope our work inspires the vision community to advance this crucial yet underexplored direction and to develop more powerful tools for scientific discovery.

\section*{Acknowledgements}
We sincerely thank A. Ren and L. Ma for their assistance with the TPL experiments. 
We gratefully acknowledge Edward. J Team for their support of the Wukong 3D model.
We also thank N. Rong, J. Niu, and D. Qi of the Analysis Center of Agriculture, Life and Environment Sciences, Zhejiang University, and J. Huang of Carl Zeiss (Shanghai) Co., Ltd., for their technical support.
We acknowledge the assistance of Y. Chen during the SEM experiments.
We are grateful to W. S{\l}{\'o}wko for insightful discussions.
This work was partially supported by the National Natural Science Foundation of China (No. 62425209).
{
    \small
    \bibliographystyle{ieeenat_fullname}
    \bibliography{main}
}
\appendix
\clearpage
\maketitlesupplementary

\setcounter{table}{0}
\setcounter{figure}{0}
\setcounter{equation}{0}
\renewcommand{\thetable}{\thesection\arabic{table}}
\renewcommand{\thefigure}{\thesection\arabic{figure}}
\renewcommand{\theequation}{\thesection\arabic{equation}}

\section{Multi-View and Multi-Detector SEM Data Acquisition}

\begin{figure}[h] 
    \centering
    \includegraphics[width=\linewidth]{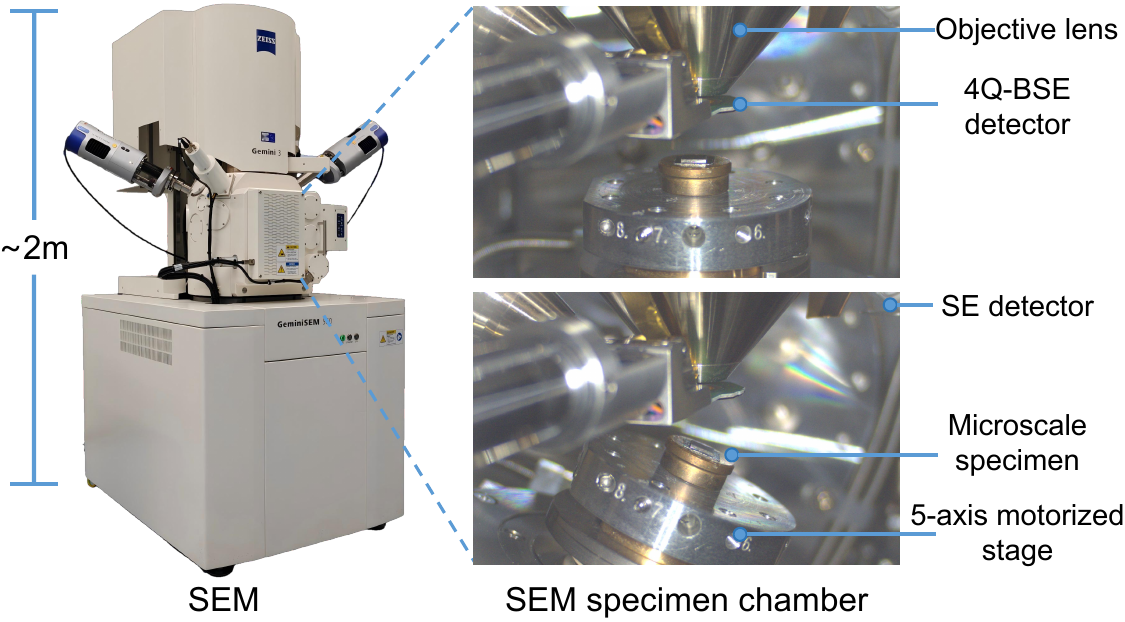}
    \caption{ \textbf{Overview of the SEM imaging setup for multi-view and multi-detector scanning.} External photograph of the SEM system used in our experiments and internal chamber views with the motorized specimen stage in flat and tilted positions.
    }
    \label{fig: SEM}
    \vspace{0em}
\end{figure}

All SEM data are acquired using a ZEISS Gemini 560 SEM system (\cref{fig: SEM}) equipped with a field-emission gun and controlled through the ZEISS SmartSEM software. Microscale specimens are mounted on a 5-axis motorized stage and sequentially rotated under SmartSEM control to enable multi-view imaging. For each microstructure, the stage is tilted along two orthogonal axes in $5^{\circ}$ increments from $-45^{\circ}$ to $+45^{\circ}$, yielding 37 distinct viewing directions.
At each viewpoint, an SE image is captured using the Everhart--Thornley detector located at the side of the SEM chamber. In addition, a pneumatically retractable 4Q-BSE detector positioned beneath the objective lens captures four BSE images from its four outer quadrants. All images are recorded at a resolution of $1024 \times 768$ pixels.
To ensure stable imaging and prevent mechanical interference during stage tilting, the accelerating voltage is set between 5--7\,kV and the working distance is maintained between 7--12\,mm. Before data acquisition, SEM parameters are optimized for each microstructure to improve image clarity and minimize charging artifacts. After optimization, all SEM parameters, including accelerating voltage, working distance, magnification, brightness, and contrast, are kept fixed throughout the entire imaging sequence.

To improve the efficiency of multi-view and multi-detector data acquisition, we utilize the compucentric function in SmartSEM, enabling the SEM to maintain focus on the microstructure as the stage tilts or rotates. 
To enable this function, the rotation center and its distance from the sample surface are determined in accordance with SmartSEM operational guidelines. 
With this information, the software automatically computes stage offsets after each stage motion, ensuring the microstructure remains centered in the field of view.
Furthermore, we packaged all data acquisition processes into a unified automated macro script, utilizing the SmartSEM macro command interface to realize multi-detector imaging and stage movements.
Therefore, operator intervention is minimal to occasional manual corrections for accumulated stage drift, resulting in a low manual workload for the entire data acquisition process.

\section{Neural Field Implementation Details}

The learnable parameters $\Theta$ of our neural field consist of both the multi-resolution hash table parameters and the MLP weights. At each training iteration, we randomly sample 256 rays, with 1024 points along each ray. Each 3D point is first encoded using multi-resolution hashing~\cite{instant_ngp}, a technique that significantly accelerates training and inference of neural fields. We employ a 16-level hash table with a 2-dimensional feature per level, producing a 32-dimensional concatenated feature vector.
This hash-encoded feature is then concatenated with the raw 3D coordinate and fed into a shallow MLP consisting of a single hidden layer with 64 units and ReLU activation.
The MLP maps the 3D coordinate to its corresponding SDF value.
This neural field representation is built upon an open-source PyTorch~\cite{pytorch} implementation~\cite{instant-nsr-pl} of NeuS~\cite{Neus}.
The SDF is subsequently converted into a density value using an unbiased and occlusion-aware weighting function~\cite{Neus}. 
Finally, volume rendering is performed along each camera ray to compute surface intersection depths $\widehat{z}$ and normals $\widehat{n}$.
For loss weighting, we set $\lambda_1 = 0.5$ for TPL samples and $0.1$ for pollen and SiC samples. The remaining weights $\lambda_2$, $\lambda_3$, and $\lambda_4$ are fixed to 0.1, 1.0, and 1.0 for all samples. The parameter $\alpha$ is set to 0.25 for TPL microstructures and 0.5 for pollen and SiC. Following prior findings on the validity range of the BSE signal model~\cite{PS31, PS47}, the BSE loss $\mathcal{L}_{\text{BSE}}$ is applied to regions with incidence angle $\theta$ less than 60 degrees.
All experiments are conducted on a workstation equipped with an AMD 5950X CPU, NVIDIA RTX 4090 GPU, and 128\,GB RAM. 
After training, we discretize the space into a 512 $\times$ 512 $\times$ 512 voxel grid and infer the SDF value at each vertex. 
These SDF values are then converted into meshes using the Marching Cubes algorithm~\cite{Marching_cubes_98} to produce 3D models shown in our experiments.

\section{Sample Preparation}
\subsection{TPL Microstructure Fabrication}
\label{sec:TPL}
Microstructures were fabricated using the commercial photoresist IP-S (Nanoscribe GmbH), which was dispensed onto an indium tin oxide-coated glass substrate (25 mm $\times$ 25 mm $\times$ 0.7 mm, Nanoscribe GmbH). A commercial direct laser writing system (Photonic Professional GT2, Nanoscribe GmbH), equipped with a 780 nm femtosecond laser (80 MHz repetition rate, 80–100 fs pulse duration) and a 25$\times$ objective lens with a numerical aperture (NA) of 0.8, was employed for the 3D printing of microstructures. CAD models were processed in Describe v2.7 (Nanoscribe GmbH) to generate executable job files, with the slicing and hatching distances set to 500 nm and 400 nm, respectively. We set printing parameters with a laser power of 50 mW and a scanning speed of 50,000 {\textmu}m/s. These job files were then transferred to NanoWrite v1.8 (Nanoscribe GmbH) to initiate the writing process. After printing, the microstructures were developed in propylene glycol methyl ether acetate (PGMEA) for 20 minutes, followed by rinsing in isopropyl alcohol (IPA) for 5 minutes at room temperature to remove unpolymerized photoresist.
Prior to SEM imaging, the printed microstructures on the substrate were sputter-coated with a thin platinum layer using an ion sputter coater (GVC-2000, Gewei Instrument) at 15 mA for 60 seconds.

\subsection{Peach Pollen and SiC Sample Preparation}
The Okubo peach pollen was sourced from Dangshan County Orchard Farm (China). The micro-scale SiC powder was purchased from Beesley New Materials (Suzhou) Co., Ltd. (China). It was produced via a high-temperature reaction between silica and petroleum coke, followed by crushing and classification to obtain particles with an average diameter of approximately 45 {\textmu}m. Both samples were prepared using the same procedure. They were mounted on 9.5 mm aluminum pin stubs using carbon conductive adhesive tabs and coated with gold using an E-1010 ion sputter coater (Hitachi) at a 15 mA current for 60 seconds, effectively preventing charging artifacts during SEM imaging.

\section{Challenges in Acquiring Ground-Truth 3D Geometry for SEM Surface Reconstruction}
Quantitative evaluation of SEM 3D reconstruction remains fundamentally challenging due to the difficulty of obtaining micro–nanoscale ground-truth surface geometry. Here, we provide a brief discussion of two primary factors.

\subsection{Limitations of Micro-Nanoscale 3D Metrology Techniques}
Although a variety of high-resolution 3D metrology techniques exist, including atomic force microscopy (AFM), white-light interferometry (WLI), confocal microscopy, X-ray micro-CT, and Focused Ion Beam Scanning Electron Microscope (FIB-SEM) tomography, it remains difficult for these techniques to provide surface measurements that are both non-destructive and matched to the resolution and depth-of-field of SEM. Each technique has intrinsic limitations: AFM offers a limited field of view and suffers from tip--sample convolution~\cite{AFM_image_artifacts}; WLI~\cite{WLI_Limitation} and confocal microscopy~\cite{Confocal_microscopy} lack SEM-level lateral resolution and are incompatible with steep or high-aspect-ratio microstructures; micro-CT provides insufficient surface fidelity~\cite{Micro_CT}; and FIB-SEM damages the sample surface during scanning~\cite{FIB_SEM}. Consequently, obtaining a reference 3D surface that faithfully matches SEM-observed geometry is highly challenging. 

\subsection{Why CAD Models Are Not Valid Ground Truth for TPL Microstructures}
Although we have CAD models used for TPL fabrication, these models cannot serve as ground truth for evaluation.
The reason is that the final printed geometry inevitably deviates from the design model due to the physics of the TPL process. As discussed in Supplementary \cref{sec:TPL}, factors such as laser power, scanning trajectory, slicing/hatching distances, voxel elongation, and development conditions all strongly influence the polymerization process and introduce fabrication-dependent variations~\cite{TPL_Optical}. 
These deviations are significant in practice. 
For instance, the layered patterns observed on the forehead of the Lucy sample (Fig.~4 in the main paper) are absent from the CAD model. They originate from the layer-by-layer characteristics of the TPL printing process rather than the intended geometry.
Consequently, using the CAD model as ground truth would incorrectly penalize reconstructions for accurately recovering real, physically printed surface features that do not appear in the ideal design model.

\section{Construction of Simulated Dataset}

\begin{figure}[h] 
    \centering
    \includegraphics[width=\linewidth]{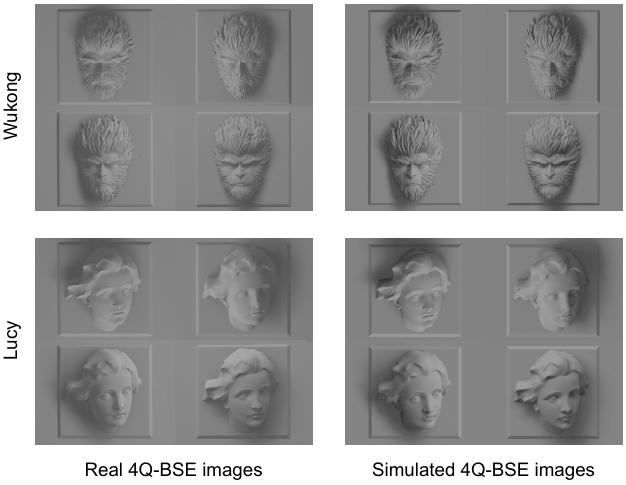}
    \caption{ \textbf{Comparison between real and simulated 4Q-BSE images.} The simulated quadrant responses closely match the real detector signals, validating both the accuracy of our learned BSE forward model and the correctness of the BSE signal formulation used in our simulation pipeline.
    }
    \label{fig: More_Sim}
\end{figure}

We construct a simulated dataset for quantitative evaluation based on the CAD models used for TPL fabrication. To simulate the multi-view SEM imaging setup, cameras are placed in Blender v4.1 at the same tilt and rotation angles as in our real experiments. Surface blur and noise are applied to the CAD models using Blender modifiers to generate coarse geometric initializations.
To approximate the appearance of real 4Q-BSE images, we simulate each component of the BSE formation process separately. Shadow-free BSE intensities are computed using our forward model $\mathcal{F}(\bar{n};\bar{\Phi})$, where $\bar{n}$ denotes the surface normals of the CAD geometry and $\bar{\Phi}$ is the set of BSE parameters estimated from NFH-SEM reconstructions of real TPL samples.
A key element of BSE imaging is the soft shadowing caused by self-occlusion. Since BSE shadow formation resembles a rendering configuration in which the four detector quadrants behave as area light sources and the electron beam serves as the view direction, we model shadows in Blender by placing quadrant-shaped area lights in fixed poses relative to the camera. The resulting shadow intensity maps $\bar{\psi}$ are rendered using the Blender Cycles engine.
To match the noise characteristics of real BSE images, we estimate the pixel-wise standard deviation ($\sigma = 0.9142$ in our case) from a flat region of a TPL microstructure and use it to synthesize additive Gaussian noise $\mathcal{N}(0,\sigma^{2})$. The final simulated BSE images are given by: 
\begin{equation}
b' = \mathcal{F}(\bar{n}; \bar\Phi) -\bar\psi +\mathcal{N}(0, \sigma^2).
\end{equation}
Finally, by combining the shadow-free intensity maps, the rendered soft shadow maps, and the synthesized noise, we obtain simulated 4Q-BSE images $b'$ with high consistency with real-world images (\cref{fig: More_Sim}).

\section{Evaluation Metrics}
We use the Chamfer distance and the average angular difference of surface normals between the reconstructed geometry and the ground truth to evaluate geometric accuracy. In addition to these standard metrics, we define two evaluation metrics for the BSE forward model and the estimated 4Q-BSE shadow maps produced by NFH-SEM.

We evaluate the accuracy of the BSE forward model $\mathcal{F}^{\widehat{\Phi}}$ with learned parameters $\widehat{\Phi}$, by comparing it to the ground-truth model $\mathcal{F}^{\bar\Phi}$ used to generate the simulated dataset:
\begin{equation}
\mathcal{E}_{\text{BSE}} = \frac{1}{4T}\sum_{i}^{A,B,C,D}\sum_{q}^{T} \left| \mathcal{F}_i(\theta_q,\varphi_i; \bar\Phi)  - \mathcal{F}_i(\theta_q,\varphi_i; \widehat{\Phi}) \right|.
\end{equation}  
By fixing the input $\varphi$ to the quadrant directions $\varphi_i$, each quadrant’s forward model depends solely on $\theta$.
We uniformly sample $T$ angles $\theta_q$ within the supervision range of $\mathcal{L}_{\text{BSE}}$, and compute the intensity difference between $\mathcal{F}^{\widehat{\Phi}}$ and $\mathcal{F}^{\bar{\Phi}}$. 
The mean error $\mathcal{E}_{\text{BSE}}$ measures how accurately the BSE forward model is learned.

Here, we aim to define a metric to demonstrate the percentage of BSE shadows successfully separated by NFH-SEM.
BSE shadows are inherently soft and do not exhibit clear, binary boundaries, making it unsuitable to evaluate shadow estimation accuracy using hard shadow masks. Moreover, stronger shadows result in a larger BSE loss and therefore have a more significant negative impact on the learning of both geometry and the BSE forward model. For these reasons, the shadow evaluation metric must consider not only the spatial distribution of shadows but also their intensity.
After training, shadow-free 4Q-BSE images are computed using the estimated surface normals $\widehat{n}$ and the learned model $\mathcal{F}^{\widehat{\Phi}}$. Subtracting the measured BSE images from these predictions yields the estimated shadow maps, $\widehat{\psi} = \left| \mathcal{F}(\widehat{n}; \widehat{\Phi}) - b'\right|$.
These shadows are not absorbed into the neural field geometry or the forward model, demonstrating NFH-SEM’s ability to disentangle shadows in BSE images.
Given the ground-truth shadow intensities $\bar{\psi}$, we define a 4Q-BSE soft shadow accuracy metric:
\begin{equation}
\mathcal{S}_{\text{shadow}} =
100 \times 
\left(
1 -
\frac{1}{4P}
\sum_{i}^{A,B,C,D}
\sum_{u}^{P}
\frac{
\sum_{v}^{Q} |\bar{\psi}_{iuv} - \widehat{\psi}_{iuv}|
}{
\sum_{v}^{Q} (\bar{\psi}_{iuv} + \widehat{\psi}_{iuv})
}
\right),
\end{equation}
where $P$ is the number of viewpoints with index $u$, and $Q$ is the number of pixels per image with index $v$.
This percentage score is computed by measuring the pixel-wise intensity difference between the estimated and ground-truth shadow maps and normalizing it by their total intensity. 
Higher values of $\mathcal{S}_{\text{shadow}}$ indicate more accurate shadow estimation.

\section{Additional Experiments}
\subsection{Results on Learning-Based MVS Methods}
We evaluate two representative learning-based MVS methods, GeoMVSNet~\cite{geomvsnet} and MVSFormer++~\cite{mvsformer++}, by applying their DTU~\cite{DTU_dataset} pretrained models to our NFH-SEM dataset. As shown in~\cref{fig: MVS_Initialization}, both methods generalize poorly to the microscopic SEM domain. Their reconstructed surfaces are highly noisy, contain large structural inconsistencies, and often fail to produce complete geometry. These observations further highlight the domain gap between conventional RGB-based MVS training data and SEM imaging.

\subsection{Robustness to Geometric Initialization}
\begin{figure*}[h]
    \centering
    \includegraphics[width=0.8\linewidth]{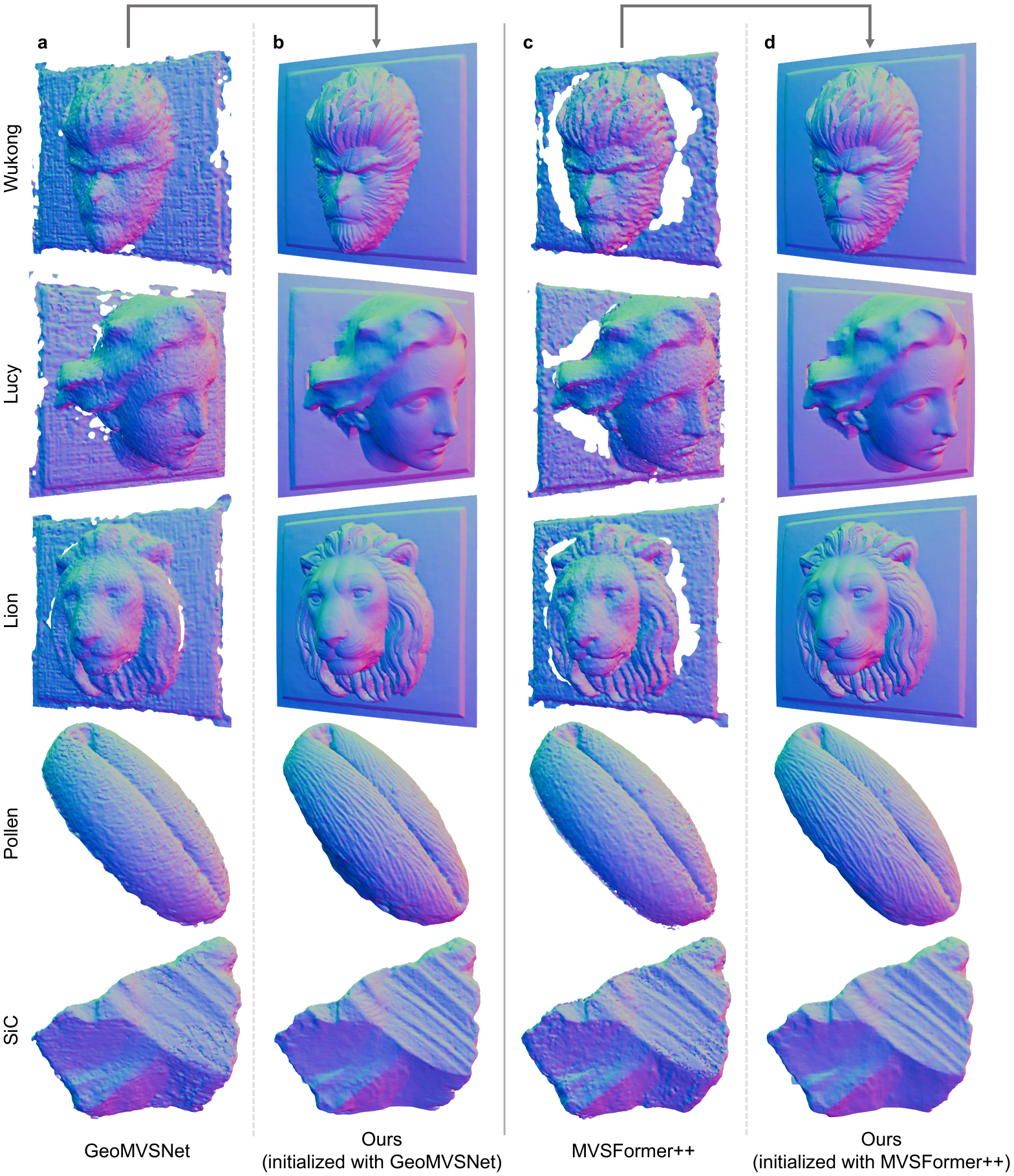}
    \vspace{0em}
    \caption[MVS_Initialization supp.]{ 
    \textbf{Evaluation of learning-based MVS methods and robustness of NFH-SEM to different initializations.}
    \textbf{(a, c)} Surface reconstructions of GeoMVSNet~\cite{geomvsnet} and MVSFormer++~\cite{mvsformer++}.
    \textbf{(b, d)} NFH-SEM reconstructions initialized with (a) and (c), respectively.
}
    \label{fig: MVS_Initialization}
\end{figure*}

NFH-SEM relies on a coarse initial geometry to guide neural field optimization. To evaluate the robustness of our method to the quality of this initialization, we replace the initialization used in the main paper with the much worse surface reconstructions generated by GeoMVSNet~\cite{geomvsnet} and MVSFormer++~\cite{mvsformer++}. As shown in~\cref{fig: MVS_Initialization}, NFH-SEM still converges to clean, complete, and high-fidelity surfaces even when initialized with geometry that contains severe noise, large missing regions, and surface holes. This experiment highlights the strong robustness of NFH-SEM to noisy or incomplete initial geometry and demonstrates the effectiveness of our reconstruction pipeline.

\subsection{Ablations on TPL Microstructures}
We provide additional qualitative ablation studies on real TPL microstructures using the same ablation settings and component definitions as in the simulation experiments presented in the main paper. As shown in~\cref{fig: supp_tpl_bse}, different strategies for gradient extraction from 4Q-BSE signals lead to notably different reconstruction results. Simplified BSE forward models or gradients derived from single-view approximations produce degraded surface geometry. As illustrated in~\cref{fig: supp_tpl_shadow}, the shadow separation strategy further improves reconstruction quality, especially around structural edges affected by strong shadowing. The qualitative trends observed on real TPL data are consistent with the quantitative results from simulations, reinforcing the importance and effectiveness of each module in NFH-SEM.

\subsection{Experiments on Simulated 3Q-BSE Data}
Most modern SEM systems are equipped with multi-quadrant BSE detectors. 
The most common configuration is a four-quadrant BSE (4Q-BSE) detector with 90$^{\circ}$ spacing between adjacent quadrants, as in the ZEISS Gemini 560 SEM used in our experiments.
In contrast, SEM systems from Thermo Fisher Scientific (formerly FEI) often employ a three-quadrant BSE (3Q-BSE) detector~\cite{PS_3Q}, which uses 120$^{\circ}$ spacing between quadrants.
While prior methods are typically tailored to a specific detector configuration, most commonly the widely used 4Q-BSE detector~\cite{PS14, PS41, PS47}, our method generalizes across different detector setups and can be readily adapted to a wide range of SEM systems.
To demonstrate the flexibility of NFH-SEM, we constructed a simulated 3Q-BSE dataset, as shown in ~\cref{fig: 3QBSE_sim}. 
All other experimental conditions remain identical to those used in our 4Q-BSE simulations, except for the detector configuration.
As shown in~\cref{fig: 3QBSE_result}, NFH-SEM reliably reconstructs accurate surface geometry under the 3Q-BSE configuration, demonstrating its flexibility and practical applicability across different SEM systems.

\begin{figure*}[h]
    \centering
    \includegraphics[width=0.88\linewidth]{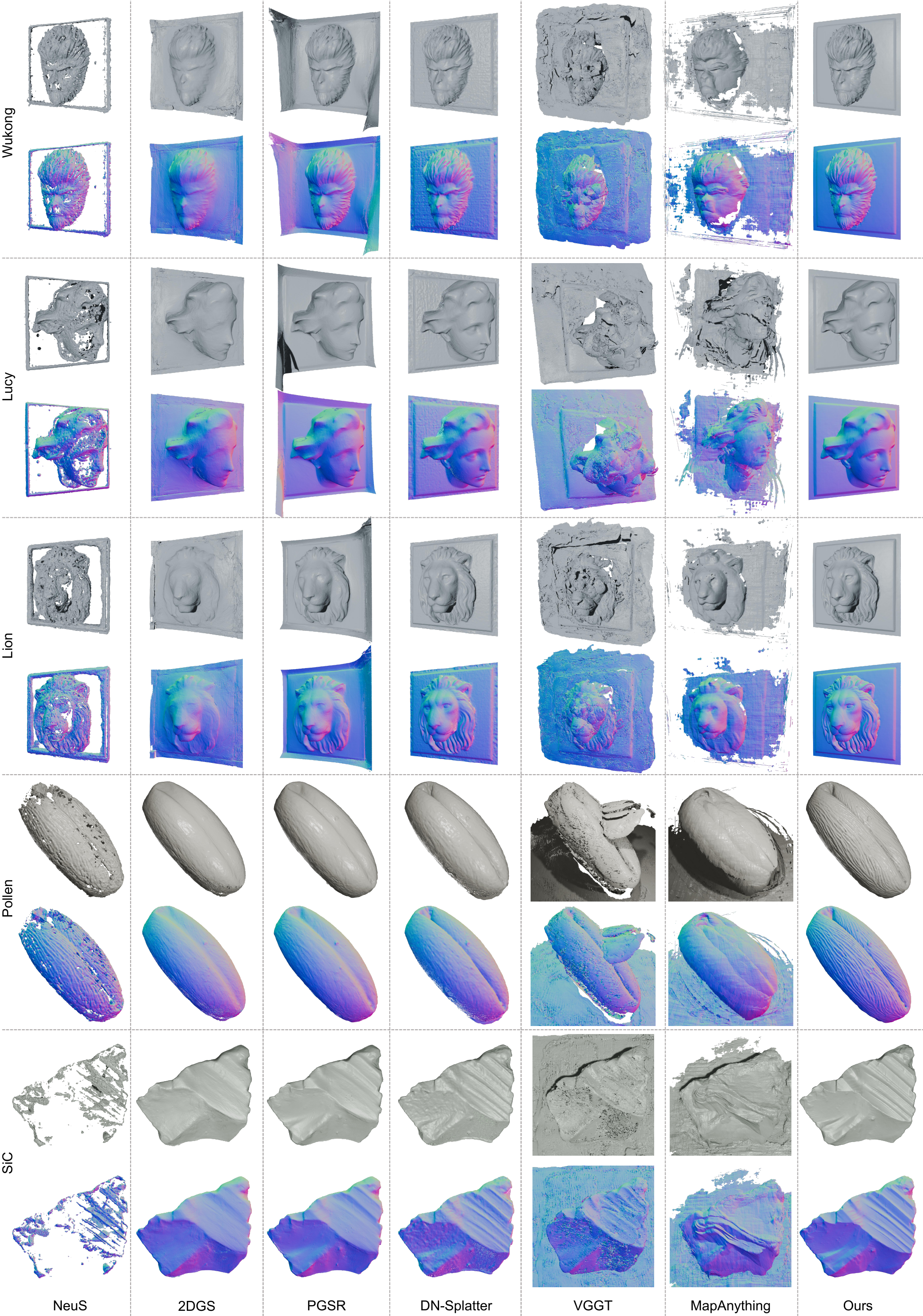}
    \vspace{-1em}
    \caption{ 
    \textbf{Results of learning-based surface reconstruction methods on the NFH-SEM dataset.}
}
    \label{fig: supp_more_learning}
\end{figure*}

\begin{figure*}[h]
    \centering
    \includegraphics[width=\linewidth]{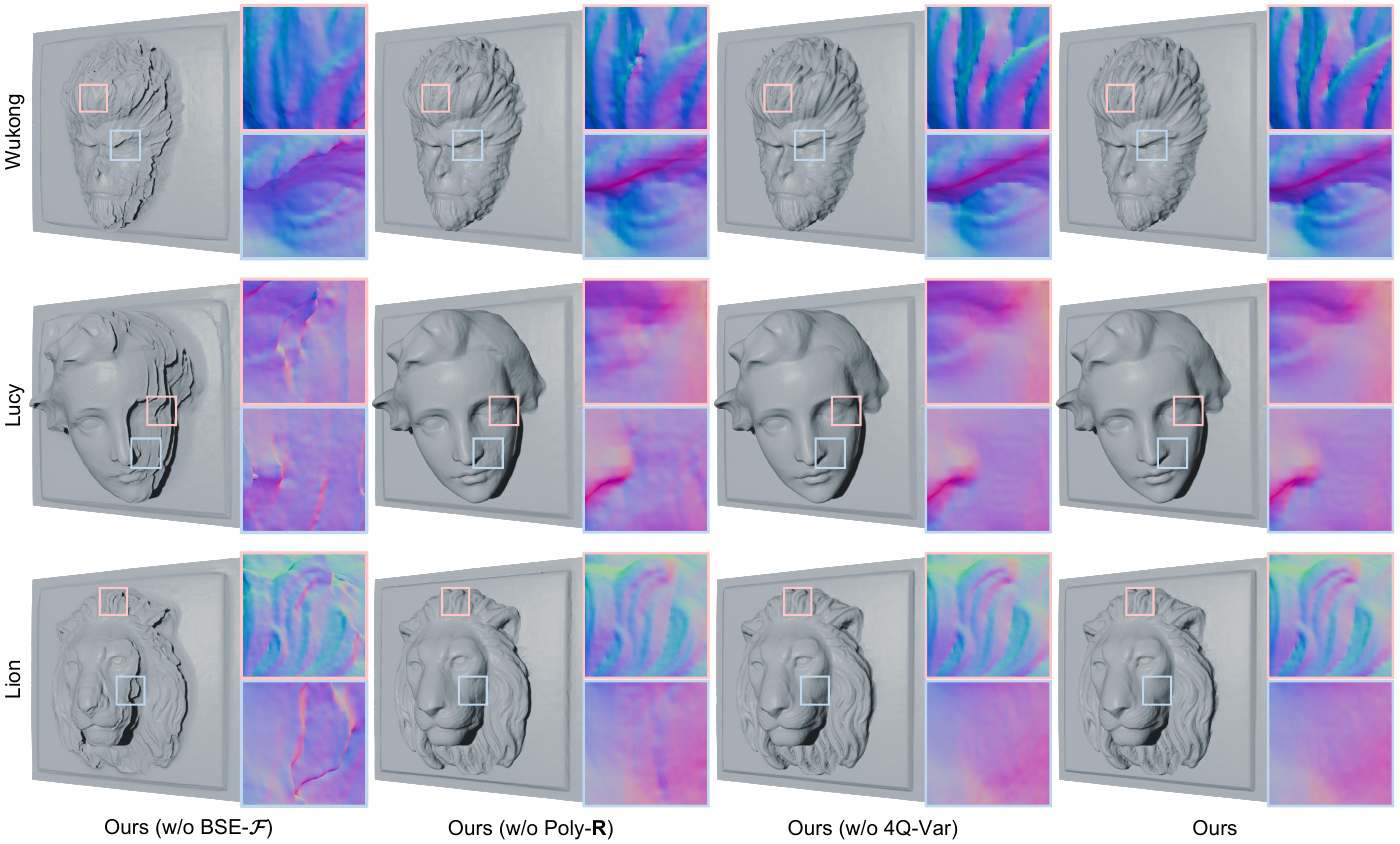}
    \caption[Qualitative comparison of different BSE forward model design on the TPL microstructures.]{ 
    \textbf{Comparison of BSE gradient extraction strategies for TPL microstructure reconstruction.} 
}
    \label{fig: supp_tpl_bse}
\end{figure*}

\begin{figure*}[h]
    \centering
    \includegraphics[width=0.9\linewidth]{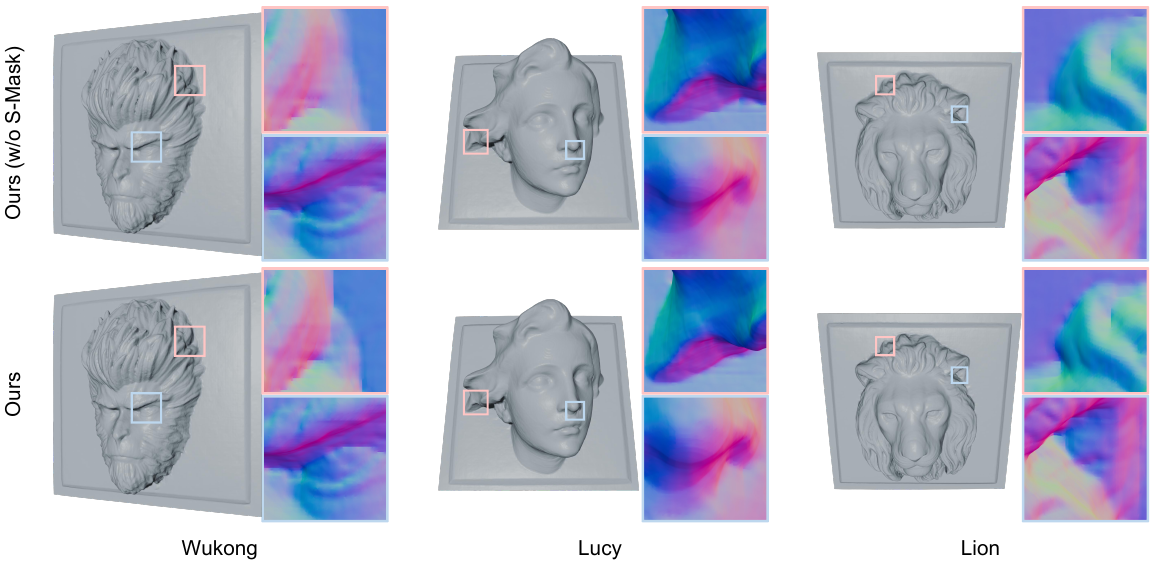}
    \caption{ 
    \textbf{Effect of the shadow separation strategy on TPL microstructure reconstruction.}
}
    \label{fig: supp_tpl_shadow}
\end{figure*}

\begin{figure*}[h]
    \centering
    \includegraphics[width=\linewidth]{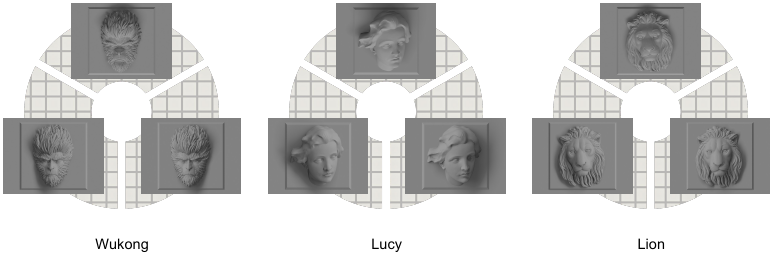}
    \caption{ 
    \textbf{Simulated BSE images acquired using a three-quadrant BSE detector.} 
}
    \label{fig: 3QBSE_sim}
\end{figure*}

\begin{figure*}[h]
    \centering
    \includegraphics[width=\linewidth]{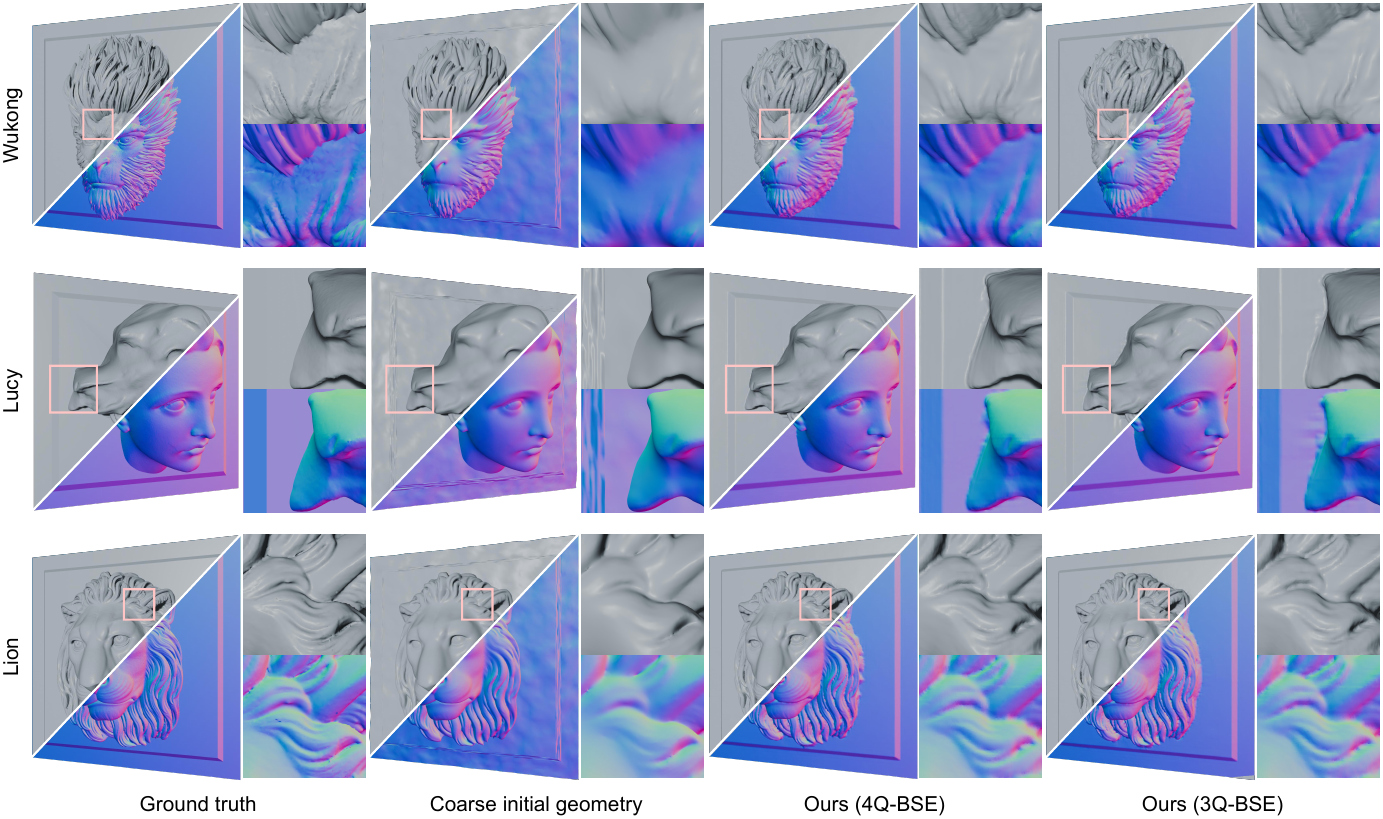}
    \caption{ 
    \textbf{NFH-SEM reconstruction results on simulated data under different BSE detector configurations.} 
}
    \label{fig: 3QBSE_result}
\end{figure*} 

\end{document}